\documentclass[%
reprint,
superscriptaddress,
amsmath,amssymb,
aps,
]{revtex4-1}

\usepackage{graphicx}
\usepackage{dcolumn}
\usepackage{bm}

\begin{document}

\preprint{APS/123-QED}

\title{Green-Kubo relation for friction at liquid/solid interface}

\author{Kai Huang}
\affiliation{%
Materials Science Program, University of Wisconsin, Madison, Wisconsin 53706-1595, USA
}
\author{Izabela Szlufarska}%
\email{szlufarska@wisc.edu}
\affiliation{%
Materials Science Program, University of Wisconsin, Madison, Wisconsin 53706-1595, USA
}
\affiliation{%
Department of Materials Science and Engineering, University of Wisconsin, Madison, Wisconsin 53706-1595, USA
}


\date{\today}

\begin{abstract}
We have developed a Green-Kubo (GK) relation that enables accurate calculations of friction at solid/liquid interfaces directly from equilibrium molecular dynamics (MD) simulations and that provides a pathway to bypass the time scale limitations of typical non-equilibrium MD simulations. The theory has been validated for a number of different of interfaces and it is demonstrated that the liquid/solid slip is an intrinsic property of an interface. Because of the high numerical efficiency of our method, it can be used in design of interfaces for applications in aqueous environments, such as nano- and micro-fluidics.

\begin{description}
\item[PACS numbers]
05.40.-a, 47.10.A-, 68.08.De
\end{description}
\end{abstract}

\pacs{Valid PACS appear here} 
\maketitle

\section{Introduction}
\begin{figure}
\includegraphics[scale =0.7]{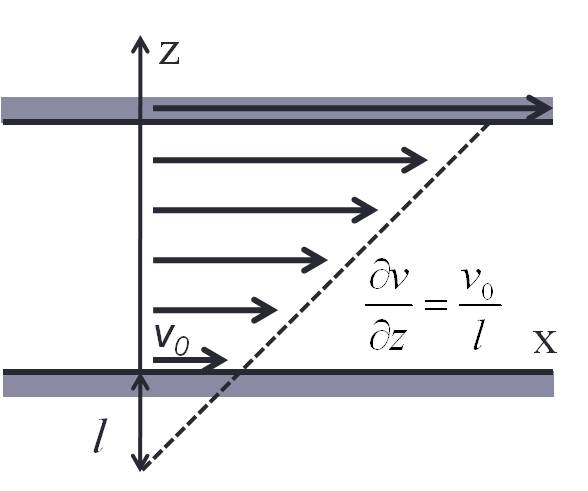}
\caption{\label{definition} Schematic representation of slip-boundary conditions. $l$ is the slip length and $v_0$ is the velocity of the liquid layer adjacent to the solid surface. }
\end{figure}

The nature of the liquid/solid (L/S) boundary conditions have been a subject of an intense scientific debate for over a century~\cite{Navier, questions, slip-review}. It is only recently that the existence of a slip at such interfaces has been accepted~\cite{confine2013, intrinsic, limits-no-slip, slip-evidence}. The urgency of understanding slip and related phenomena has increased further with the miniaturization of devices. In particular, in micro- and nano-fluidics~\cite{nanofluidics,transport-nanofluidics,whitesides} the presence or absence of slip and the magnitude of friction at the L/S interface will have a large effect on the flow rate of the fluid. L/S slip is often characterized by a slip length $l$ or a friction coefficient
\begin{equation}
\bar{\eta}=\eta/l, 
\label{sliplength}
\end{equation}
where $\eta$ is the viscosity of the liquid. Slip length is defined as the extrapolated distance where the velocity of the liquid matches the velocity of the solid wall, as shown in Fig.~\ref{definition}. One of the major challenges in this field is the ability to measure or predict the slip length, or alternatively the coefficient of friction at the L/S interface. While experimental measurements are plagued with their own limitations (see for instance~\cite{CNT}), here we focus on the challenges associated with predicting the friction coefficient from atomistic simulations. In particular, non-equilibrium molecular dynamics (NEMD) simulations, which can be invaluable in providing insights into relations between interfacial properties and friction~\cite{MM,NEMD,Troian,Robbins}, are limited (with few exceptions, see for instance Ref.~\cite{viscosity}) to sliding velocities and shear rates that are orders of magnitude higher than in typical experiments. In the case of L/S friction, such simulations often trigger a non-linear behavior, leading to qualitative deviations from experiments. More explicitly, for the viscous friction law 
\begin{equation}
F=-\bar{\eta}u,
\label{LSF}
\end{equation}
where $F$ is the friction force per unit area, $u$ is the slip velocity, the non-linear behavior means that at large enough $u$, $\bar{\eta}$ is not constant but it depends on $u$. Here, we use the linear response theory and the Generalized Langevin Equation (GLE) to derive a new relationship for predicting friction from equilibrium molecular dynamics (EMD) simulations. Because friction is calculated from equilibrium properties of the system, by nature this approach overcomes the time scale limitations of traditional NEMD techniques. Our new theory is validated by comparing results of EMD simulations with NEMD results extrapolated to the limit of low sliding velocities.

A pioneering attempt to predict friction for L/S interfaces from EMD simulations was reported by Bocquet and Barrat (BB)~\cite{BBGK}. It was proposed that the coefficient of friction $\bar{\eta}$ can be calculated from the integral of the time correlation function of the total friction force $F_{\mathrm{tot}}$ between the solid and the liquid layer adjacent to the solid
\begin{equation}
\bar{\eta}=\frac{1}{SkT}\int_0^\infty \langle F_{\mathrm{tot}}(0)F_{\mathrm{tot}}(t) \rangle_{\mathrm{EC}}\,\mathrm{d}t.
\label{BBGK}
\end{equation}
In the above expression, $S$ is the surface area, $k$ is the Boltzmann constant, $T$ is temperature and $\langle ... \rangle_{\mathrm{EC}}$ denotes ensemble averages at equilibrium condition. 
Application of the above theory led to two numerical issues, recognized by the authors themselves~\cite{BBGK}: 1) Eq.~(\ref{BBGK}) vanishes for finite systems and a cut-off time in the upper limit of the integral is necessary to predict the integral in the thermodynamic limit. 2) Predictions of Eq.~(\ref{BBGK}) do not agree quantitatively with the fitted parameters based on the transverse momentum density autocorrelation function.  
In addition, a number of criticisms have been raised in literature, questioning whether $\bar{\eta}$ in Eq.~(\ref{BBGK}) corresponds to the intrinsic properties of the interface~\cite{comment,comment2,comment3}. BB responded to these criticisms in another paper~\cite{BBGK2} and explained that results from simulations could be spurious if the limit of the fluid particles going to infinity (the thermodynamic limit) and the limit of time going to infinity are not taken in the proper order. In this debate  one critical issue was ignored, which is that friction coefficient is not a bulk property. Therefore, a large volume of the liquid is not necessary for friction to arise at the L/S interface. For example, L/S friction is present in a nanotube~\cite{water-nt,nanotube1,thermal-nano} where the number of liquid molecules is highly constrained and in such confined systems~\cite{confine2013,leng} one is not allowed to take the thermodynamic limit of the system size. Furthermore, L/S friction is local, which means that for an inhomogeneous solid surface~\cite{patterned}, friction coefficient can be different from one domain to another, and for a mixture of liquid~\cite{binary}, friction coefficient can be different from one kind of liquid particle to another. It is straightforward to see that due to its mathematical structure, expressions like Eq.~(\ref{BBGK}) cannot capture the potential inhomogeneity of L/S friction.

In this paper, we first develop a formal GK relation for L/S friction that overcomes the limitations of previous models and allows high efficient numerical evaluation of friction. Subsequently, we validate our GK relation numerically by demonstrating a very good agreement between the predictions from our GK relation for the L/S friction coefficient and the measurements from NEMD simulations. Finally, for completeness we compare performance of the newly developed GK relation and that proposed previously by BB. 

\section{Theoretical model}
\subsection{General strategy for the derivation of a Green-Kubo relation for L/S friction}

Having recognized that L/S friction is local and shall not be described by a bulk-like transport coefficient that vanishes with finite system size, we shall construct the GK relation from the dynamics of individual liquid particles near the solid wall. Our approach to deriving a GK relation therefore differs from the standard one summarized by McQuarrie~\cite{McQuarrie}, which starts from the Fourier transform of the diffusion-like partial differential equation
\begin{equation}
\frac{\partial\phi}{\partial t}=D\nabla^2\phi,
\label{diff}
\end{equation}
where $\phi$ is the field of interest and $D$ is the corresponding transport coefficient.
This difference in the derivation strategies arises from the difference between the L/S friction coefficient and thermal transport coefficients, such as viscosity and thermal conductivity. Due to the discontinuity at the L/S interface, the L/S friction described by Eq.~(\ref{LSF}) does not have a form of a partial differential equation in Eq.~(\ref{diff}) that describes thermal processes at the macroscale. In fact, L/S friction is in general a mechanical process rather than a thermal one, although sometimes it can be strongly coupled to thermal processes in the system~\cite{thermal-jump,arun,thermal-slip,thermal-nano}. The mechanical nature of L/S friction makes it possible for us to construct a mechanical external Hamiltonian and apply the linear response theory. Here, in order to directly take into account the microscopic details of the L/S interface, we choose to apply the external perturbation to an individual liquid particle at the L/S interface. The linear response theory then allows us to find out the expression for $\bar{\eta}_i$, which is the friction coefficient of an individual liquid particle $i$ near the solid interface. Finally, we can sum the contributions from all the interfacial particles to obtain the total friction coefficient
\begin{equation}
\bar{\eta}=\frac{1}{C}\sum_i\eta_i,
\label{sum}
\end{equation}
where $C$ is the normalization factor. For a flat surface, $C$ can be chosen as the unit surface area. The specific choice of $C$ is not as important. What is most important is that Eq.~(\ref{sum}) demonstrates the additive property of L/S friction, which allows for modeling of inhomogeneous L/S interfaces.

\subsection{Application of linear response theory}
As we will apply the linear response theory twice in our derivation of the GK relation, we shall first briefly review the linear response technique. When a system at thermal equilibrium is slightly perturbed by an external force $f$, the response of the system can be predicted from the time correlation function of its thermal fluctuations at the equilibrium state.
For any physical observable $B$ of interest, its thermal average at the perturbed non-equilibrium state can be expressed as the convolution of the external force and the generalized susceptibility $\chi_{AB}$, as follows 
\begin{equation}
\Delta\langle B(t) \rangle=\int_{-\infty}^\infty \chi_{AB}(t-t')f(t')\mathrm{d}t'.
\label{LRT}
\end{equation}
Here, $A$ is the internal variable that is conjugate to $f$. $\Delta\langle B(t) \rangle=\langle B(t) \rangle_{\mathrm{NE}}-\langle B \rangle_{\mathrm{EC}}$, where $\langle ... \rangle_{\mathrm{NE}}$ and $\langle ... \rangle_{\mathrm{EC}}$ denote ensemble averages at non-equilibrium and equilibrium conditions, respectively. Due to the causality, $\chi_{AB}$ is non-zero only at finite times (the application of the external force $f$ begins at time zero). The relation between the susceptibility $\chi_{AB}$ and the corresponding time correlation of $\delta A$ and $\delta B$ at equilibrium is as follows
\begin{equation}
\chi_{AB}(t)= \begin{cases} -\frac{1}{kT}\frac{d}{dt}\langle \delta A(0)\delta B(t) \rangle_{\mathrm{EC}}, & t\ge0 \\ 0, & t<0 \end{cases}
\label{chi}
\end{equation}
where $\delta A=A-\langle A \rangle_{\mathrm{EC}}$ and $\delta B=B-\langle B \rangle_{\mathrm{EC}}$ are thermal fluctuations in variables $A$ and $B$, respectively.

For the L/S interface, we choose the perturbation Hamiltonian to be $\Delta H=-xf\mathrm{e}^{\mathrm{i}\omega t}$, where $f\mathrm{e}^{\mathrm{i}\omega t}$ is the external drag force, $x$ is the particle's displacement along the direction parallel to the solid wall, $\omega$ is frequency, and $t$ is time. One can thus obtain the Fourier transformed frequency dependent susceptibility by applying the periodic external force. 
Under the perturbation of $\Delta H$, the liquid particle will respond with drift velocity, the magnitude of which is determined by the balance between the external drag force, the friction force exerted by the solid wall, and the friction force exerted by the surrounding liquid.
We first choose $A=x_i$ and the physical observable of interest $B=u_i$, where $u_i$ is the drift velocity of the interfacial particle $i$ within a plane parallel to the solid wall. By substituting Eq.~(\ref{chi}) into Eq.~(\ref{LRT}) and taking a Fourier transform, one can show that $u_i$ is proportional to the velocity autocorrelation function (or the mobility $\mu$) determined in the equilibrium system 
\begin{equation}
\langle u_i \rangle_\omega(t)=\frac{f\mathrm{e}^{\mathrm{i}\omega t}}{k T}\int_0^\infty \langle u_i(0)u_i(t) \rangle_{\mathrm{EC}} \mathrm{e}^{-\mathrm{i}\omega t}\,\mathrm{d}t,
\label{slipv}
\end{equation}
\begin{equation}
\mu_i(\omega)=\frac{1}{kT}\int_0^\infty \langle u_i(0)u_i(t) \rangle_{\mathrm{EC}} \mathrm{e}^{-\mathrm{i}\omega t}\,\mathrm{d}t.
\label{mu}
\end{equation}
In the next step, we choose $B$ to be the friction force $F_i$ exerted by the solid wall on a single interfacial particle $i$ while retaining $A=x_i$. One can then show that $F_i$ is related to the correlation between the particle's velocity and the friction force experienced by the particle at equilibrium
\begin{equation}
\langle F_i \rangle_\omega(t)=\frac{f\mathrm{e}^{\mathrm{i}\omega t}}{kT}\int_0^\infty \langle u_i(0)F_i(t) \rangle_{\mathrm{EC}} \mathrm{e}^{-\mathrm{i}\omega t}\,\mathrm{d}t.
\label{LRF}
\end{equation}
By definition, the friction coefficient $\bar{\eta}_i$ is equal to the ratio between the friction force and the slip velocity. Using Eq.~(\ref{slipv}) and Eq.~(\ref{LRF}) we can write
\begin{equation}
\bar{\eta}_i(\omega)\equiv -\frac{\langle F_i \rangle_\omega(t)}{\langle u_i \rangle_\omega(t)}=-\frac{\int_0^\infty \langle u_i(0)F_i(t) \rangle_{\mathrm{EC}} \mathrm{e}^{-\mathrm{i}\omega t}\,\mathrm{d}t}{\int_0^\infty \langle u_i(0)u_i(t) \rangle_{\mathrm{EC}} \mathrm{e}^{-\mathrm{i}\omega t}\,\mathrm{d}t}.
\label{etai}
\end{equation}
We can now sum up the microscopic friction coefficients $\bar{\eta}_i$ in Eq.~(\ref{etai}) and normalize the sum by the area $S$ of the interface to obtain the macroscopic friction coefficient
\begin{equation}
\bar{\eta}(\omega)=-\frac{1}{SkT\mu_i(\omega)}\sum_i\int_0^\infty \langle u_i(0)F_i(t) \rangle_{\mathrm{EC}} \mathrm{e}^{-\mathrm{i}\omega t}\,\mathrm{d}t.
\label{eta}
\end{equation}
The order of the sum and the integral can be switched without affecting the results.
It should be noted here that the sum in Eq.~(\ref{eta}) can be taken over all the liquid particles since the liquid particles away from the interfacial region have no contribution to the integral due to the short-range nature of friction force ($F_i=0$). The friction force between the liquid and the solid is either intrinsically short-range (as in the case of hydrogen or covalent bonding) or it is screened by water (for electrostatic interactions). So far we assumed that there is only one type of particle in the liquid. It is straightforward to generalize Eq.~(\ref{eta}) to a mixture of liquids, $\mathrm{A}, \mathrm{B}, \mathrm{C}, ...$, based on the additive property of the friction coefficient shown in Eq.~(\ref{sum}). One can simply use the same method to evalute the friction coefficient for different types of particles seperately and then sum them up to get $\bar{\eta}=\bar{\eta}_{\mathrm{A}}+\bar{\eta}_{\mathrm{B}}+\bar{\eta}_{\mathrm{C}}+...$. For instance, to calculate the friction coefficient for particles of type A, one first needs to determine mobility $\mu_{\mathrm{A}}$ using Eq.~\ref{mu}, plug it into Eq.~(\ref{eta}) and take the summation in Eq.~(\ref{eta}) over all liquid particles of type A ($\sum_{i\in\mathrm{A}}$).

\subsection{Reformulation using the Generalized Langevin Equation (GLE)}
Equation~(\ref{eta}) shows that the friction coefficient is inversely proportional to the liquid interfacial mobility $\mu$. However, unfortunately this equation is not particularly practical for simulations because $\mu$ needs to be calculated for particles at the L/S interface. Liquid particles are free to diffuse away from the interface and it turns out that the finite amount of time a particle spends near the interface is not necessarily sufficient to obtain a well-converged estimate for $\mu$. In addition, the evaluation of the interfacial mobility could be sensitive to the definition of the interfacial region. The uncertainty in the width of the interface will be transfered to the uncertainty in $\bar{\eta}$. Lastly, to obtain a reliable  $\bar{\eta}$, one will need to repeat the calculation for various interfacial widths to find a best fit or average. To avoid the above issue, we will rewrite Eq.~(\ref{eta}) using the Generalized Langevin Equation (GLE) formalism~\cite{Zwanzig,Snook}. The GLE generalizes the Brownian motion by taking into account the memory of the particle, which means that the friction force experienced by a liquid particle depends on the history of the particle's motion~\cite{BM1,BM2}. Because we are interested in calculating the L/S friction force $F_i$, in our formulation of GLE $F_i$ is represented explicitly instead of being adsorbed into the random force $R_i$ and/or the memory kernel $\gamma_i$. Thus, the extended new GLE reads
\begin{equation}
m_i\dot{u}_i(t)=-\int_0^t\gamma_i(t-t')u_i(t')\,\mathrm{d}t'+R_i(t)+F_i(t),
\label{GLE}
\end{equation}
where $m_i$ is the mass of the particle $i$, $\gamma_i$ stands for the memory kernel and $R_i$ represents the random force. 
We assume the following three properties that are associated with the GLE 
\begin{equation}
\langle u(0)R(t) \rangle_i=0, \;t>0,
\label{A1}
\end{equation}
\begin{equation}
\langle F(0)R(t) \rangle_i=0, \;t>0,
\label{A2}
\end{equation}
\begin{equation}
\langle R(0)R(t) \rangle_i=kT\gamma_i(t), \;t>0.
\label{A3}
\end{equation}
Because $u$ and $F$ are antisymmetric and symmetric, respectively, under time reversal, the correlation between them is antisymmetric under time reversal, which leads to $\langle u(0)F(t) \rangle_i=-\langle F(0)u(t) \rangle_i$. 

Now we can use the GLE to derive a relation between different time correlation functions.
It is straightforward to show that any two of the three properties of the GLE above (Eqs.~(\ref{A1}-\ref{A3})) can lead to the third one. Here, we start with the GLE (Eq.~\ref{GLE}) and Eqs.~(\ref{A1}) and (\ref{A2}). 
For simplicity of the expression, we introduce the following abbreviation 
\begin{equation}
\langle AB \rangle(\omega) \equiv \int_0^\infty \langle A(0)B(t) \rangle_{\mathrm{EC}} \mathrm{e}^{-\mathrm{i}\omega t}\,\mathrm{d}t.
\end{equation}
By applying $\langle A...\rangle$ with $A$ being $u,R,F,\dot{u}$ to both the left-hand-side (LHS) and the right-hand-side (RHS) of the GLE (Eq.~(\ref{GLE})), we obtain the following four equations, respectively
\begin{equation}
\langle uF \rangle_i(\omega)=[\mathrm{i}\omega m_i+\gamma_i(\omega)]\langle uu \rangle_i(\omega)-kT,
\label{u}
\end{equation}
\begin{equation}
[\mathrm{i}\omega m_i+\gamma_i(\omega)]\langle Ru \rangle_i(\omega)=\langle RR \rangle_i(\omega)+\langle RF \rangle_i(\omega),
\label{R}
\end{equation}
\begin{equation}
\langle FF \rangle_i(\omega)=[\mathrm{i}\omega m_i+\gamma_i(\omega)]\langle Fu \rangle_i(\omega),
\label{F}
\end{equation}
\begin{equation}
\begin{split}
m_i\langle \dot{u}\dot{u}\rangle_i(\omega)=\gamma_i(\omega)\langle u\dot{u} \rangle_i(\omega)+\langle RR \rangle_i(\omega) \\
+\langle FR \rangle_i(\omega)-\mathrm{i}\omega \langle uF \rangle_i(\omega).
\label{udot}
\end{split}
\end{equation}
To derive the above equations, we used $\langle \dot{u}u \rangle_i(\omega)=-\langle u\dot{u} \rangle_i(\omega)=\langle u^2 \rangle_i-\mathrm{i}\omega\langle uu \rangle_i(\omega)$, $\langle \dot{u}\dot{u} \rangle_i(\omega)=\mathrm{i}\omega\langle \dot{u}u \rangle_i(\omega)$ and $\langle \dot{u}F \rangle_i(\omega)=-\mathrm{i}\omega\langle uF \rangle_i(\omega)$. With Eqs.~(\ref{u}-\ref{udot}) and $\langle u(0)F(t) \rangle_i=-\langle F(0)u(t) \rangle_i$, one can show the following
\begin{equation}
\langle Fu \rangle_i(\omega)=kT-[\mathrm{i}\omega m_i+\gamma_i(\omega)]\langle uu \rangle_i(\omega),
\label{Fu}
\end{equation}
\begin{equation}
\langle RR \rangle_i(\omega)=m_i\gamma_i(\omega)\langle u^2 \rangle_i=kT\gamma_i(\omega),
\label{RR}
\end{equation}
\begin{equation}
\langle Ru \rangle_i(\omega)=\gamma_i(\omega)\langle uu \rangle_i(\omega),
\label{Ru}
\end{equation}
\begin{equation}
\langle RF \rangle_i(\omega)=\gamma_i(\omega)\langle uF \rangle_i(\omega).
\label{RF}
\end{equation}
Having derived a close set of the relations between various time correlation functions, we will briefly comment on some of them. First of all, Eq.~(\ref{Ru}) can be rewritten as $\gamma_i(\omega)=\langle Ru \rangle_i(\omega)/\langle uu \rangle_i(\omega)$. This expression is a counterpart of Eq.~(\ref{etai}), which described the friction coefficient of particle $i$, where friction originates from the surrounding liquid. Equation~(\ref{Fu}) can be rewritten as $kT/\langle uu \rangle_i(\omega)=\mathrm{i}\omega m_i+\bar{\eta}_i(\omega)+\gamma_i(\omega)$, which simply means that the total friction coefficient (LHS of the above equation) is the sum of inertia (first term on the RHS of the above equation), the L/S friction coefficient (second term on the RHS of the above equation) and the liquid/liquid (L/L) friction coefficient (third term on the RHS). Typically, for the case of slip boundary conditions, we expect $|\bar{\eta}_i(\omega)|\ll |\gamma_i(\omega)|$. Finally, one should note that Eq.~(\ref{RR}) is just the Fourier transform of Eq.~(\ref{A3}) and Eq.~(\ref{RR}) shows that the fluctuation-dissipation relation is the result of the lack of correlation of the random force $R$ (Eqs.~(\ref{A1},\ref{A2})) to the velocity and friction force.


We are now ready to derive the final expression for $\bar{\eta}_i$ and $\bar{\eta}$. As Eq.~(\ref{u}) relates the L/S friction force-velocity correlation to the mobility of the liquid particle and Eq.~(\ref{F}) connects the L/S friction force autocorrelation function to the L/S friction force-velocity correlation, we can rewrite Eq.~(\ref{etai}) and Eq.~(\ref{eta}) as
\begin{equation}
\bar{\eta}_i(\omega)=\frac{\langle FF \rangle_i(\omega)}{kT-\langle Fu \rangle_i(\omega)},
\label{etai2}
\end{equation}
\begin{equation}
\bar{\eta}(\omega)=\frac{\sum_i\langle FF \rangle_i(\omega)}{SkT(1-\alpha(\omega))},
\label{eta2}
\end{equation}
where $\alpha(\omega)\equiv{\langle Fu \rangle_i(\omega)}/kT$. At zero frequency $\omega=0$, one can show from Eq.~\ref{Fu} that $\bar{\eta}_i(0)/\gamma_i(0)=\alpha(0)/(1-\alpha(0))$. For slip boundary conditions, this ratio is expected to be very small, leading to $\alpha(0)\ll 1$, which will be shown later to be true in our simulations. Equation~(\ref{eta2}) is the new Green-Kubo (GK) relation for the macroscopic coefficient of friction coefficient that does not require calculation of the interfacial mobility and can be directly evaluated from EMD simulations. 
To numerically evaluate Eq.~(\ref{eta2}) at $\omega=0$, the only parameter one needs to choose by hand is the number density $n$ of interfacial liquid particles, as there is no clear boundary of the L/S interface. Since $n$ enters Eq.~(\ref{eta2}) only through $\alpha(0)=\sum_i\langle Fu \rangle_i(0)/kTSn$, the uncertainty in Eq.~(\ref{eta2}) from $n$ will be suppressed by the fact that $\alpha\ll 1$ or $\bar{\eta}_i\ll\gamma_i$ for a slip boundary condition.

$\bar{\eta}_i$ and $\bar{\eta}$ are in general complex numbers for a finite frequency $\omega$ and they become real numbers for $\omega=0$.
It is worth pointing out the macroscopic friction coefficient obtained in this way is not limited to a certain geometry (e.g., the curvature of area $S$ is not required to be zero) since it is not calculated from macroscopic correlation defined on the area $S$, but from the microscopic correlation that are not dependent on the global geometry. This property allows our method to be applicable to curved interfaces, such as surfaces of nanotubes or even nanoparticles~\cite{JR}.

The GLE formalism, used to derive Eq.~(\ref{eta2}), merits a few additional comments. First of all, the GLE written in the form of Eq.~(\ref{GLE}) has many applications. For example, it has been utilized to explain the diffusion of impurities and defects in crystals~\cite{defects}, the superionic conductance~\cite{GLE3} and the fluctuations of the Josephson supercurrent through a tunneling junction~\cite{GLE1,GLE2}. Similar GLE has also been implemented in the Brownian dynamics simulations~\cite{BD}. The physical meaning of the last term of Eq.~(\ref{GLE}), $F(t)$ varies from case to case. It is important to note that in most of the applications, the memory function $\gamma(t)$ is approximated as a delta function $\gamma\delta(t-t')$, where $\gamma$ is a constant, not a function of time anymore. Such treatment of coarse graining the memory function constitutes a compromise between a mathematical rigor and practicality of the applications (which requires simple form of the memory function for computational efficiency), because the exact form of the memory function is often difficult or impossible to obtain. Nevertheless, in our case one does not need to know the exact form of the memory function in order to be able to reformulate Eq.~(\ref{etai}) into Eq.~(\ref{etai2}). Therefore, irrespectively of the exact form the memory function, our GK relation for the coefficient of friction expressed in Eq.~(\ref{eta2}) is formal and exact as long as the GLE given in Eq.~(\ref{GLE}) can formally describe the motion of an interfacial liquid particle. In fact, one can formally construct Eq.~(\ref{GLE}) for $\vec{F}=-\nabla U(\vec{r})$ (where $U$ is an arbitrary external potential) using the projection operator approach~\cite{Zwanzig,Snook,Berne}. Proof of this statement is given in the Appendix. In the derivation of the extended GLE, we found that the memory kernel and the friction coefficient are in general two dimensional tensors. This is not surprising, since due to the perturbation of the solid wall, the memory and the transport coefficient of the interfacial particles could be anisotropic~\cite{aniso1, aniso2, aniso3}. As a result, Eq.~(\ref{GLE}) is generalized as
\begin{equation}
m_i\dot{\vec{u}}_i(t)=-\int_0^t\boldsymbol{\gamma}_i(t-t')\vec{u}_i(t')\,\mathrm{d}t'+\vec{R}_i(t)+\vec{F}_i(t),
\label{GGLE}
\end{equation}
where $\dot{\vec{u}}_i, \vec{u}_i, \vec{R}_i$ and $\vec{F}_i$ are two dimensional vectors that lie parallel to the solid wall and $\boldsymbol{\gamma}_i$ is the tensorial memory function with the generalized fluctuation-dissipation relation $kT\boldsymbol{\gamma}_i(t)=\langle \vec{R}_i(0)\vec{R}_i(t)\rangle$. In the same spirit, Eqs.~({\ref{A1}, \ref{A2}}) are generalized to $\langle \vec{u}_i(0)\vec{R}_i(t)\rangle=\langle \vec{F}_i(0)\vec{R}_i(t)\rangle=0$ at a finite time $t$. Consequently, the friction coefficient becomes a tensor and the reformulation of GK relation reads
\begin{equation}
\begin{split}
\boldsymbol{\bar{\eta}}_i(\omega)&=[kT\boldsymbol{\mu}_i(\omega)]^{-1}\langle \vec{F}\vec{u} \rangle_i(\omega)\\
&=\langle \vec{F}\vec{F} \rangle_i(\omega)[kT\mathrm{\bf{I}}-\langle \vec{F}\vec{u} \rangle_i(\omega)]^{-1},
\label{Getai}
\end{split}
\end{equation}
where $\boldsymbol{\mu}_i(\omega)=\langle \vec{u}\vec{u} \rangle_i(\omega)/kT$ is the tensorial mobility.
If the $x$ and $y$ axes parallel to the solid wall are chosen to align with the crystallographic symmetry axes of the solid surface, then $\boldsymbol{\gamma}_i$ and $\boldsymbol{\bar{\eta}}_i$ could be diagonalized. For simplicity, we will limit the discussion in the simulation section to such a situation because we choose to adapt this particular alignment.
In addition to the formal proof in the Appendix, the validity of the GLE equation will be further tested against results of MD simulation in the next section.

\section{MD simulation results and discussions}
\subsection{Simulation test of the Generalized Langevin Equation (GLE) for interfacial liquid particles}

\begin{figure} 
\includegraphics[width=\columnwidth,viewport=10 0 730 570,clip=]{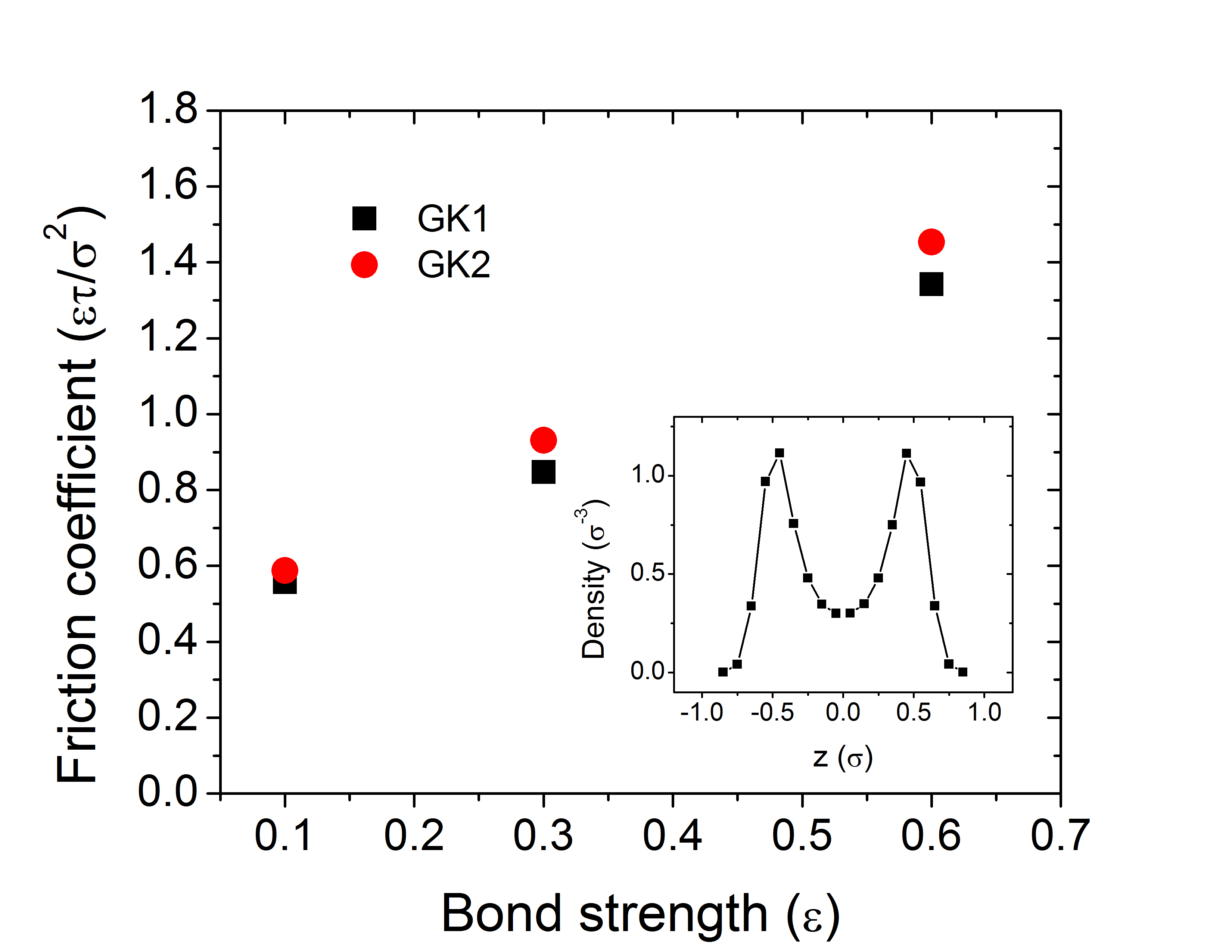}
\caption{Friction coefficients calculated from Eq.~(\ref{eta}) (the GK relation, GK1) and Eq.~(\ref{eta2}) (the reformulated GK relation by GLE, GK2) calculated as a function of bond strength between liquid and solid atoms. $\sigma$ is a reduced unit of length, as explained in the main text. Inset: Density profile of a hard-sphere liquid confined between two solid walls located at $z=\pm 1.2\sigma$.}
\label{layer}
\end{figure}

We carried out MD simulations to numerically validate our GK relation. 
First, in order to test if the extended GLE with the assumption $\langle F(0)R(t) \rangle_i=0$ captures the physics of interfacial liquid particles or not, we compare predictions of the friction coefficient from Eq.~(\ref{eta}) and Eq.~(\ref{eta2}).
One of the challenges of using Eq.~(\ref{eta}) was the calculation of local liquid mobility at the L/S interface.
To overcome this challenge, we designed a simulation system, where the liquid is confined between the two solid walls to the extent that almost the entire body of the liquid becomes interfacial (see the inset of Fig.~\ref{layer}). The solid walls are face-centered cubic crystals with a constant surface area of $48\sigma\times48\sigma$, where $\sigma$ is the unit of length in reduced Lennard Jones (LJ) units. There are 8000 hard-sphere liquid particles confined between the walls.
Using this setup, we calculated the friction coefficient from both Eq.~(\ref{eta}) and Eq.~(\ref{eta2}) and the results are shown in Fig.~\ref{layer}. The excellent agreement between the two ways to calculate friction coefficient numerically justified our application of the extended GLE with the form of Eq.~(\ref{GLE}).

\subsection{Agreement between EMD and NEMD results}

\begin{figure}
\begin{center}
\includegraphics[width=\columnwidth]{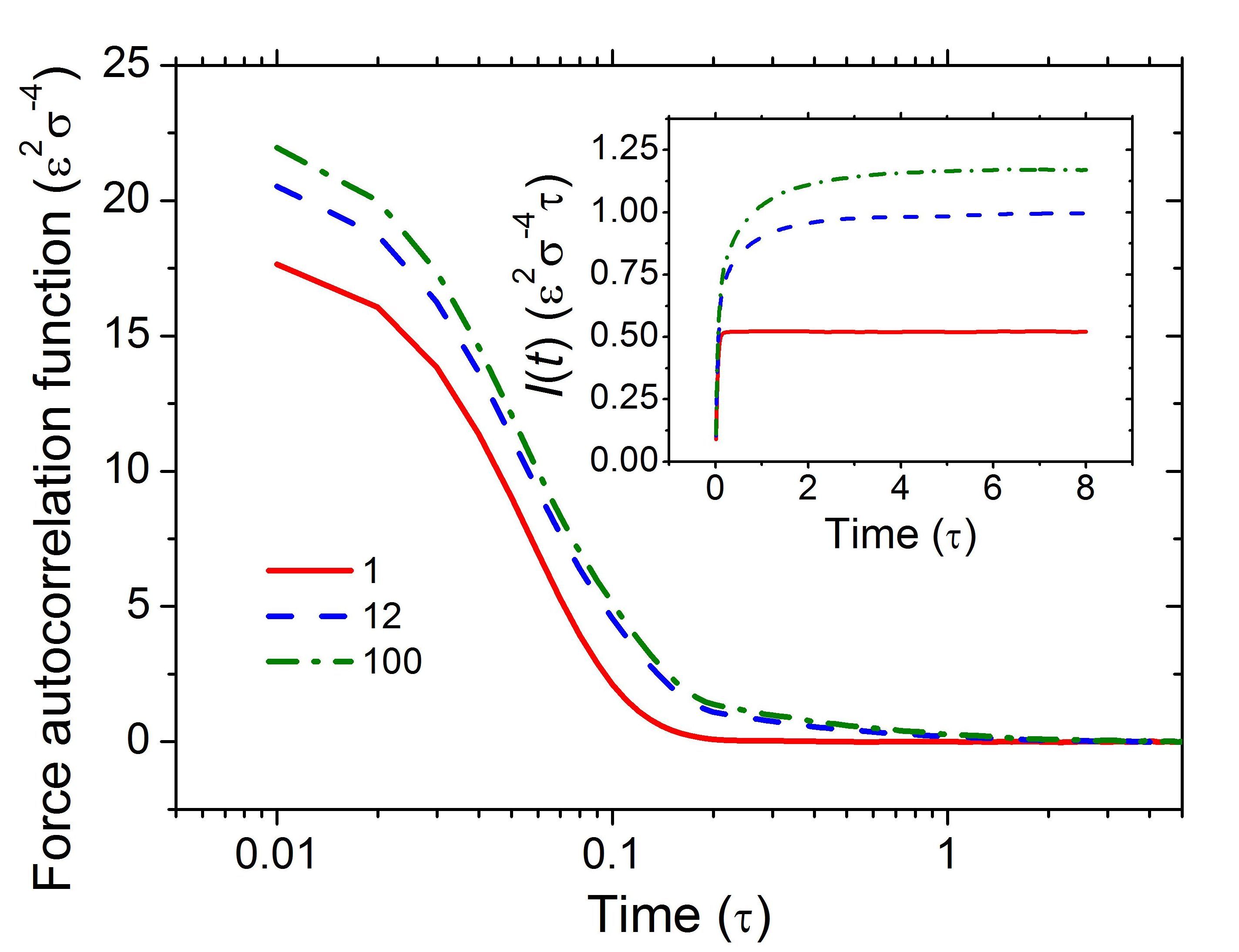}
\caption{\label{FF} (Color online) L/S frictional force autocorrelation function (FAF) and its time integral $I(t)$ (inset). Solid, dashed, and dotted-dashed lines (red, blue, and green colors online) correspond to the hard-sphere, 12-bead polymer, and 100-bead polymer liquids, respectively. Here the bond strength between liquid particles and the solid is $0.3\epsilon$.}
\end{center}
\end{figure}

\begin{figure}
\begin{center}
\includegraphics[width=\columnwidth]{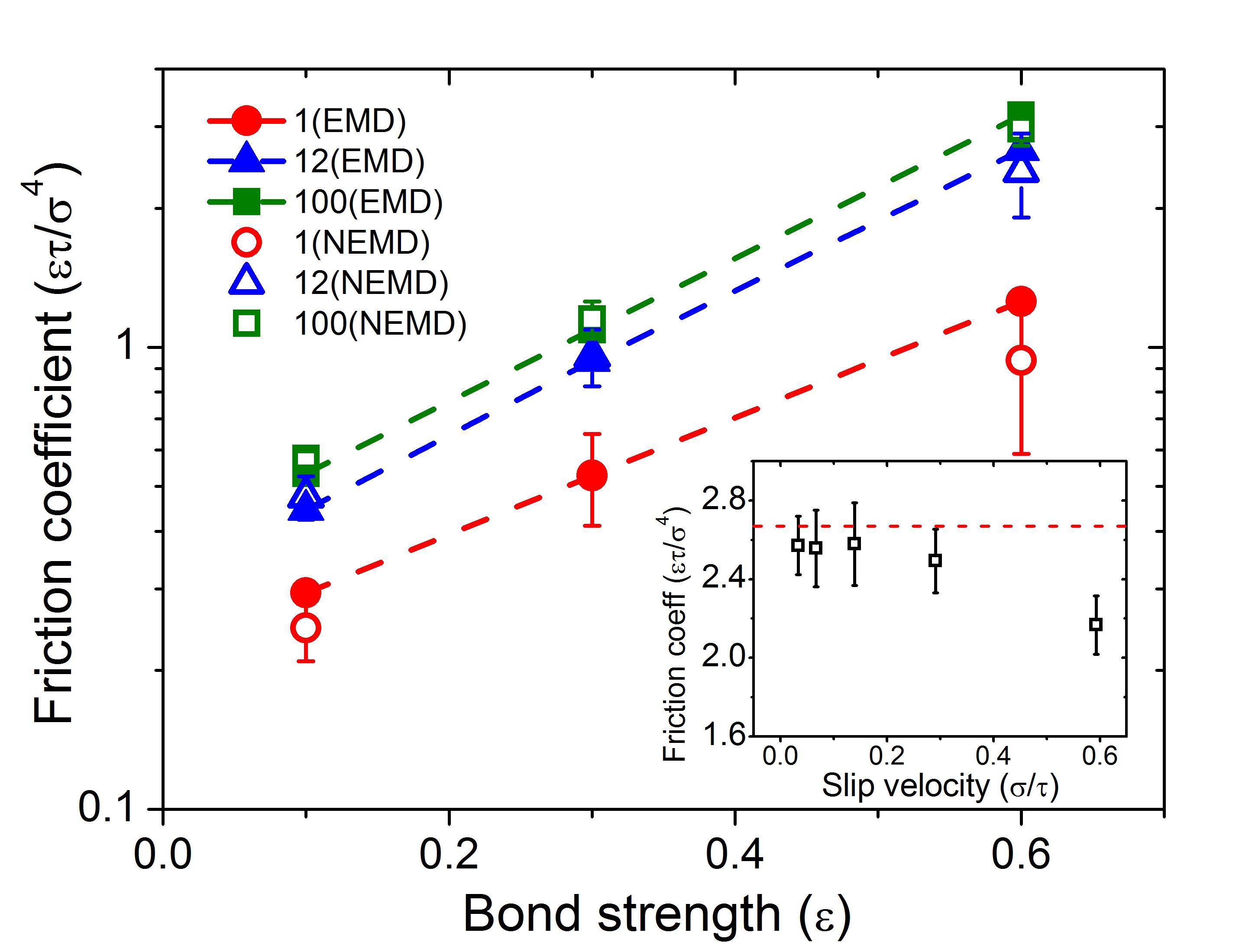}
\caption{\label{compare} (Color online) Comparison between friction coefficients at zero frequency calculated from EMD (filled symbols) and NEMD (open symbols) simulations as a function of the bond strength between the liquid particles and solid atoms. Circles, triangles, and squared (red, blue, and green colors online) in the main figure correspond to the hard-sphere, 12-bead, and 100-bead liquids, respectively. Data in the inset is calculated for interfacial bond strength of $0.6\epsilon$ for 12-bead polymer melts. The dashed horizontal line in the inset corresponds to the friction coefficient calculated from EMD simulations.}
\end{center}
\end{figure}

The next and most important examination of our proposed GK model is to see if it can predict the friction coefficient measured in a direct way, which in our case is the NEMD method. Explicitly, we calculate the friction coefficient $\bar{\eta}(\omega)$ from the EMD simulations combined with Eq.~(\ref{eta2}) and from NEMD simulations in the limit of low sliding velocities. The simulated system consists of a liquid confined between two solid walls. The simulation boxes for EMD and NEMD are identical except in EMD the solid walls are stationary, while in NEMD the walls are sliding against each other to build a shear rate in the confined liquid. To ensure that our conclusions are general, we use both, a hard-sphere liquid and a spring-bead polymer melt (for polymer liquid, index $i$ in Eq.~(\ref{eta2}) runs over the beads). We also choose two different lengths of the polymer liquid with the short one being 12 beads and the long one 100 beads per particle. The number of liquid particles in each simulation is 48000, which means 4000 molecules for the 12-bead polymer melts and 480 molecules for the 100-bead polymer melts.
The solid walls are again face-centered cubic crystals with a constant surface area of $48\sigma\times48\sigma$. The distance between the two walls is kept around $20\sigma$, varying with the liquid type to keep zero-pressure conditions. 
Periodic boundary conditions are applied within the plane of the solid wall. To explore a range of slip boundaries, we choose three different bond strengths ($0.1\epsilon, 0.3\epsilon, 0.6\epsilon$) between the solid wall and liquid particles, where $\epsilon$ is the unit of energy in LJ units. Temperature is kept at 1.1 (in LJ units) during the production run. Temperature is controlled with the Nose-Hoover thermostat coupled only to one direction, which lies within the plane of the solid surface and is perpendicular to the direction of sliding. The time step is set to be 0.002$\tau$, where $\tau=(m\sigma^2/\epsilon)^{1/2}$ and $m$ is mass of the liquid particle in LJ units. For EMD simulation, the production simulation is 5000 time steps long (10$\tau$) while the production simulation of NEMD simulation is as long as $10^6$ time steps (2000$\tau$).

It is found in simulations that $\alpha$ in Eq.~(\ref{eta2}) is generally small compared to 1, which, as discussed earlier, is expected in the case of slip boundary conditions. As a result, the autocorrelation of the L/S friction force ($\sum_i\langle F(0)F(t) \rangle_i$) is the dominant contribution to the friction coefficient $\bar{\eta}$. In Fig.~(\ref{FF}), we show the behavior of this force autocorrelation function (FAF) and its time integral ($I(t)=\frac{1}{S}{\int_0^t \sum_i\langle F_i(0)F_i(t') \rangle_{\mathrm{EC}} \,\mathrm{d}t'}$) as a function of time. The latter is important for the evaluation of friction coefficient at zero frequency limit, which is of most interest and can be compared directly to our NEMD results. For both, hard-sphere and polymeric liquids the FAF decays dramatically at short time scales and exhibits a hydrodynamic tail at longer time scales. The short time decay largely determines the growth of the time integral $I(t)$ and the hydrodynamic long tail barely contributes to the friction coefficient. One important consequence of this fast decay of FAF and the corresponding fast convergence of its time integral is that calculations of the L/S friction coefficient in EMD are two orders of magnitude faster than the NEMD calculations of the friction coefficient at a single value of the sliding velocity (simulations with multiple values of sliding velocities are needed to determine the low-velocity limit). In general, the convergence of $I(t)$ slows down as the molecular weight of the polymeric liquid increases.

In Fig.~(\ref{compare}), a comparison is made between the results from EMD and NEMD simulations. Excellent agreement is found between the friction coefficient predicted by our GK relation from EMD results and the friction coefficient calculated from NEMD in the limit of low sliding velocities (shear rates). Convergence of the NEMD simulations to the low velocity limit is illustrated in the inset of the figure. The agreement between EMD and NEMD results is found for all types of liquids considered in our study and for a range of interfacial bond strengths, which indicates that our relationship is universal. We did not show the error bars of the EMD results in Fig.~(\ref{compare}) as they are smaller than the symbol size.
The high efficiency and accuracy of the EMD method based on our GK relation enables a comprehensive exploration of the fundamentals of L/S friction, such as the dependence of the friction coefficient on pressure, wettability, surface morphology~\cite{morphology}, liquid confinement~\cite{confine2013,leng}, etc. Here, as an example, we only briefly discuss the dependence of the friction coefficient on the bond strength between liquid and solid molecules/atoms and on the properties of the liquid. From Fig.~(\ref{compare}) we can see that $\bar{\eta}$ increases roughly exponentially with the bond strength for a relatively wide range of liquids we tested. For all types of liquid we found that $\bar{\eta}$ in general increases nonlinearly with the length of the polymer chain that the liquid is made of. Interestingly, for the 12-bead and 100-bead polymer melts, the difference in $\bar{\eta}$ is very small. This trend is in consistent with the finding in Ref.~\cite{NEMD} that beyond chain lengths of about 10 beads, the molecular weight dependence of the slip length $l$ is dominated by the bulk viscosity $\eta$ (see Eq.~(\ref{sliplength}) for the relation between $\bar{\eta}$, $l$ and $\eta$).

Once the friction coefficient $\bar{\eta}$ is known, one can use it to calculate the lateral (i.e., in the plane of the solid wall) mobility $\mu$ of the interfacial liquid using Eq.~(\ref{eta}). We found in our hard-sphere simulation that the $\mu/\mu_0$ values for L/S bond strengths of 0.1$\epsilon$, 0.3$\epsilon$, 0.6$\epsilon$ are 1.72, 1.38 and 0.86, respectively, where $\mu_0$ stands for the bulk liquid mobility, which is calculated using Eq.~(\ref{mu}) in a simulation system consisting of liquid only. The observed trend of decreasing interfacial mobility with increasing L/S bond strength is not surprising, but the ability to evaluate this interfacial property can be valuable to a number of other studies, such as those focused on understanding the fundamental nature of hydrophobic interactions~\cite{Garde,Chandler,Berne2}.

\subsection{Frequency dependent L/S friction coefficient}

\begin{figure}
\includegraphics[width=\columnwidth,viewport=40 20 730 545,clip=]{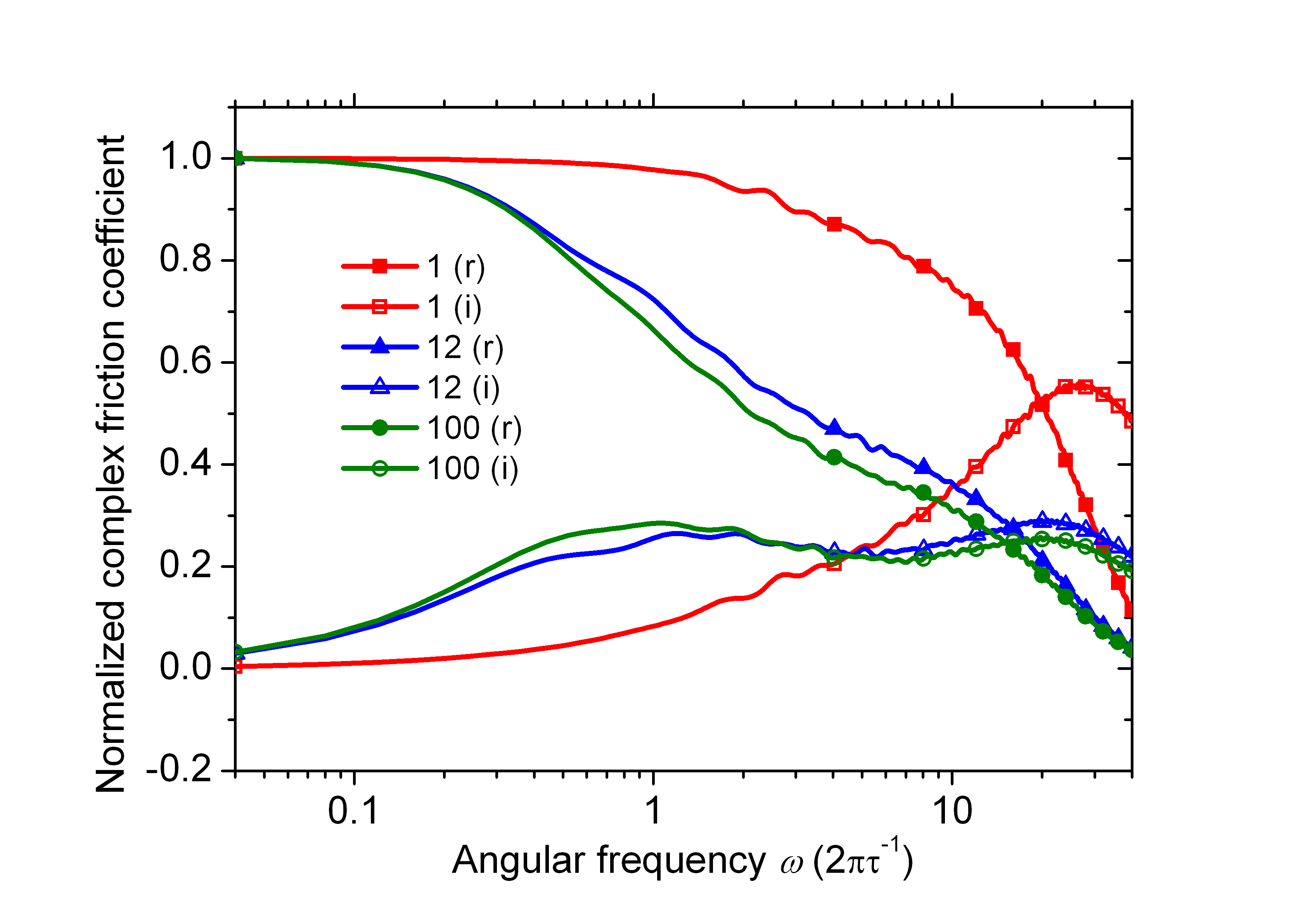}
\caption{\label{frequency} Complex friction coefficient $\bar{\eta}(\omega)$ normalized by the zero-frequency value $\bar{\eta}(0)$ as a function of frequency. Solid and dashed lines correspond to the real (r) and imaginary (i) parts of the friction coefficients. Red, blue and green colors represent hard-sphere, 12-bead, and 100-bead liquids, respectively.}
\end{figure}

\begin{figure*}
\begin{tabular}{cc}
\includegraphics[width=\columnwidth,viewport=25 0 720 560,clip=]{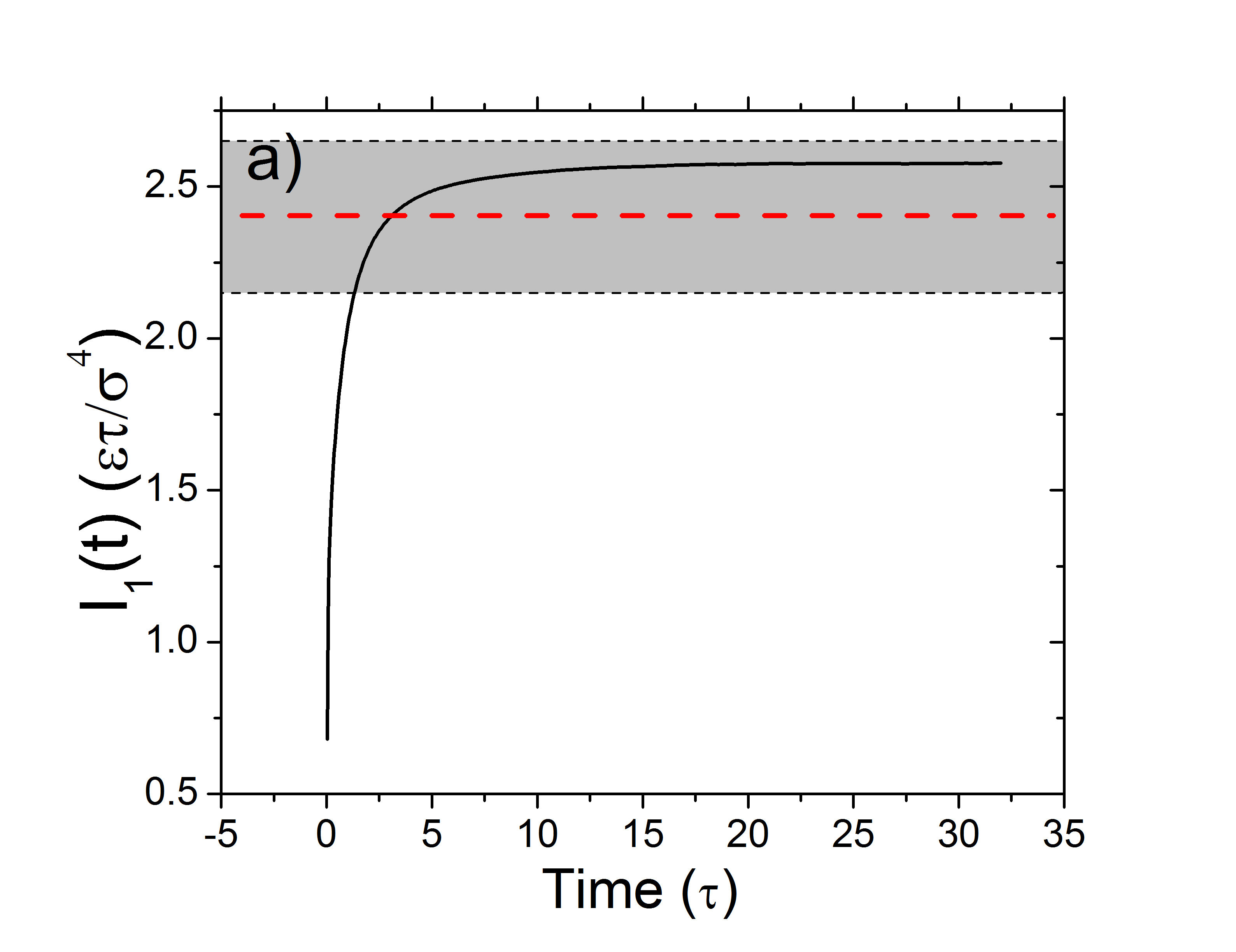}
\includegraphics[width=\columnwidth,viewport=30 0 720 560,clip=]{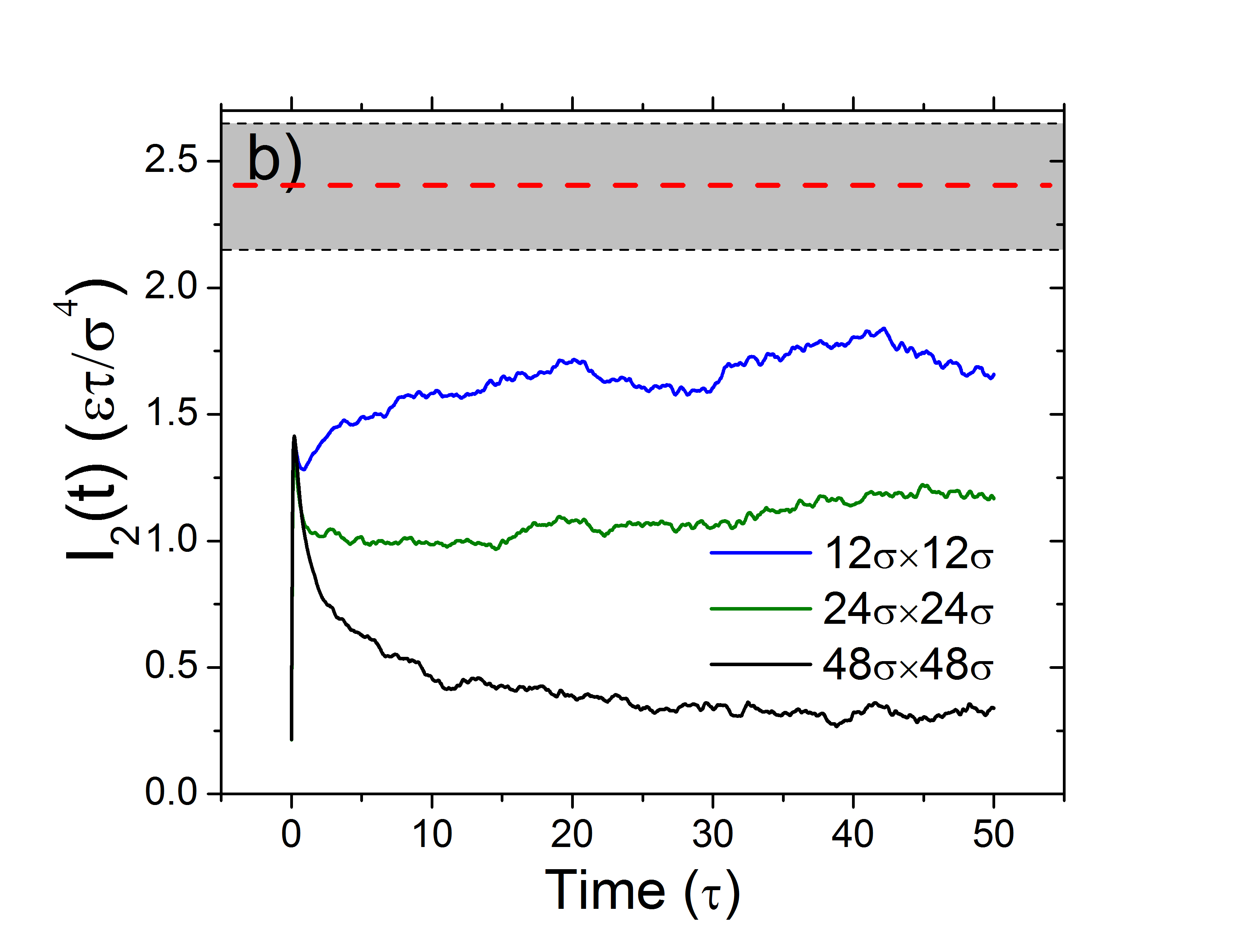}
\end{tabular}
\caption{Convergence of the time integral of friction force time correlation function (a) proposed in this work and (b) defined in Refs.~\cite{BBGK,BBGK2}. In Fig. (b), the total friction force is calculated on samples with surface areas of $12\sigma\times 12\sigma$ (blue), $24\sigma\times 24\sigma$ (green), and $48\sigma\times 48\sigma$ (black). The red dashed line represents NEMD prediction in the limit of low velocities. The height of the grey domain corresponds to the size of the error bar. Simulations were performed for a 12-bead polymer melt liqud with the bond strength between the liquid and the solid being $0.6\epsilon$. }
\label{BB}
\end{figure*} 

For any GK relation, one shall be able to get the dynamic transport coefficient from the Fourier transform of the corresponding memory function.
Here we demonstrate that our new GK relation provides access to information about dynamic properties of the L/S friction. Knowing the frequency dependent friction coefficient and mobility is of particular importance in high frequency resonators, such as those based on quartz crystal microbalance~\cite{QCM1,QCM2}. It is because at high frequencies transport coefficients can significantly deviate from their static (zero frequency) values~\cite{HK}. From Eq.~(\ref{eta2}) one can extract the frequency dependent friction coefficient by Fourier transforming the time correlation of the friction force. This coefficient connects the frequency dependent L/S friction and the slip velocity, explicitly, $F(\omega)=-\bar{\eta}(\omega)u(\omega)$. At finite frequency, $\bar{\eta}(\omega)$ is in general a complex number, meaning that there exists a phase difference between the friction and slip velocity. Figure~(\ref{frequency}) shows the details of the frequency dependence of the complex friction coefficient $\bar{\eta}(\omega)$. While the real part (solid lines) decreases monotonically, two peaks are found in the imaginary part (dashed lines). These peaks correspond to the relaxation times of the two regimes of exponential decays visible in Fig.~(\ref{FF}) (the first regime extends in time from 0.01$\tau$ to 0.1$\tau$ and the second regime from 0.1$\tau$ to 1$\tau$). In Fig.~(\ref{frequency}), the left peak, which corresponds to the slower structural relaxation, is much higher in the polymeric liquid than in the hard-sphere simple liquid. The positions of peaks in the $\bar{\eta}(\omega)$ plot (Fig.~(\ref{frequency})) depend on the properties of the liquid and these peaks could be used to design sensors based on high-frequency resonators for characterization of soft matter (especially thin films with large slip lengths).

\subsection{Comparison to the earlier Green-Kubo relation}

Lastly, it is instructive to compare the numerical performance of our model to the one reported earlier by BB in Refs~\cite{BBGK,BBGK2} and to discuss the differences and similarities between the models. 
In Fig.~\ref{BB} we plot the behavior of the time integral of two friction force autocorrelation functions (FAFs). In Fig.~\ref{BB}(a) we plot $I_1(t)=\frac{1}{SkT(1-\alpha(0))}{\int_0^t \sum_i\langle F_i(0)F_i(t') \rangle_{\mathrm{EC}} \,\mathrm{d}t'}$, which is a time integral derived in our model to predict the coefficient of friction from EMD. In Fig.~\ref{BB}(b) we show $I_2(t)=\frac{1}{SkT}{\int_0^t \langle F_{\mathrm{tot}}(0)F_{\mathrm{tot}}(t') \rangle_{\mathrm{EC}} \,\mathrm{d}t'}$, which is the equivalent time integral proposed in Ref.~\cite{BBGK,BBGK2}. The results are compared to coefficients of friction predicted in NEMD simulations in the limit of low sliding velocities (red dashed lines). One can see that predictions from EMD calculations based on our Green-Kubo relation fall within the range of error bars (grey area) of NEMD calculations. In contrast, the EMD results in Fig.~\ref{BB}(b), although on the same order of magnitude, do not agree with the NEMD results very well. In addition, we see that the disagreement becomes worse as the size of the sampling interface increases (i.e., the decay of the integral of the correlation function decays faster). This is the opposite trend than expected given the fact that as the system size approaches the thermodynamic limit (i.e., the the sampling size is increased), the time integral that defines the transport coefficient should decay slower~\cite{size1,size2,size3}.
We shall also point out that our GK relation allows a high numerical efficiency. Despite the fact that our simulations based on Eq.~(\ref{eta2}) are one order of magnitude shorter than those based on BB theory and Eq.~(\ref{BBGK}), the former approach provides a much smoother well-converged curve than the latter approach does. Specifically, data in Fig.~\ref{BB}(a) is obtained in 5,000 time steps, while it takes 50,000 time steps to obtain data in Fig.~\ref{BB}(b).

Although it is not the goal of our paper, one can speculate on the possible sources of the discrepancies between the BB model and the results of NEMD simulations. We find that there is a number of assumptions in the derivations of BB model that need to be further justified: 
(i) In the first derivation in the main text of Ref.~\cite{BBGK}, an external perturbation Hamiltonian defined with a shear rate and a reference position $z_0$ was constructed in order to apply the linear response theory and to calculate friction/slip length. As the slip length is constrained by the quantity $z_{wall}-z_0$, where $z_{wall}$ is the position of the solid wall, such a choice of external perturbation could have overspecified the problem. (ii) In the second derivation in the appendix of Ref.~\cite{BBGK}, contributions from higher (more than 2) order terms of $k$ in the normal direction to the solid wall are ignored, where $k$ is the wave vector. This truncation of higher order terms relies on the assumption that any spatial inhomogeneities are smoothly varying in the hydrodynamic (long time-scale) limit. However, due to the presence of the solid wall, properties of the liquid (e.g.,  the liquid density and viscosity) can be have pronounced inhomogeneities at the nanometer length scales along the direction normal to the wall surface. Higher order terms with respect to $k$ may be needed to capture such small inhomogeneities. Therefore the approximation of using only second-order terms in the expansion with respect to $k$ needs to be carefully justified.
(iii) In the third derivation of the BB formalism in Ref.~\cite{BBGK2}, a Brownian model is utilized to describe the motion of the solid wall. The coefficient of friction between the wall and the liquid is assumed to be equal to the integral of time correlation of the total force experienced by the solid.  From the fluctuation-dissipation theorem, one can see that such a relation is only an approximation of the formal one between the friction constant and the random force (see the force correlation section of Ref.~\cite{Kubo}). Further test is needed for the assumed approximation when replacing the random force autocorrelation function with the one of the total force.

\section{Conclusions}
In conclusion, we have developed a general GK relation for liquid/solid friction and verified its validity by numerical simulations based on the molecular dynamics technique. This new relation provides access to dynamic properties of the L/S friction and allows overcoming the challenge of limited time scales typical for NEMD simulations. We show that in the limit of low shear-rates, the coefficient of friction is not infinite (corresponding to no-slip boundary conditions), but instead it has a finite value. Consequently, the coefficient of friction is an intrinsic property of the system. Friction coefficient was also shown to be a tensor, which implies that in general it can be  anisotropic. In addition, the friction coefficient has additive properties, which means it can be calculated locally. Finally, because calculations with the new method are significantly faster (2 to 3 orders of magnitude) than traditional NEMD simulations, our GK relation opens up a new opportunity for computational exploration of L/S friction at molecular level and for rational design of L/S interfaces optimized for their slip and friction.

\begin{acknowledgments}
The authors gratefully acknowledge support from the NSF grant CMMI-0747661 and from NSF-NSEC at the UW-Madison (grant DMR-0832760).The authors gratefully acknowledge use of facilities and instrumentation supported by the University of Wisconsin Materials Research Science and Engineering Center (DMR-1121288).
\end{acknowledgments}

\appendix*
\section{Derivation of the extended GLE based on the projection operator approach}
\setcounter{figure}{0}
\setcounter{equation}{0}
\renewcommand{\theequation}{A\arabic{equation}}
\renewcommand{\thefigure}{A\arabic{figure}}

In the main text we showed that Eq.~(\ref{etai}) can be reformulated into Eq.~(\ref{etai2}) if the motion of interfacial liquid particle can be described by an extended GLE (Eq.~(\ref{GLE})). Here we construct an extended GLE using the projection operator technique~\cite{Zwanzig,Snook,Berne} for a diffusive particle moving in an external potential $U(\vec{r})$ with the resulted external force being  $\vec{F}=-\nabla U(\vec{r})$. We show that under the condition $\langle \vec{F}\rangle=0$, one can retrieve the properties in Eqs.~(\ref{A1}-\ref{A3}).


Motivated by connecting different correlation functions, we extended the original GLE by keeping the $F_i$ term out of the memory kernel and the random force. Therefore we first decompose the acceleration $\dot{\vec{u}}_i$ of an interfacial particle into to two parts: due to the external potential (the solid wall) and due to the surrounding liquid, that is
\begin{equation}
\dot{\vec{u}}_i=\mathrm{i\bf{L}}\vec{u}_i=\mathrm{i\bf{A}}\vec{u}_i+\mathrm{i\bf{B}}\vec{u}_i=\vec{F}_{\mathrm{A}}+\vec{F}_{\mathrm{B}},
\label{Force decomposition}
\end{equation}
where $\mathrm{\bf{L}}$ is the Liouville differential operator, and $\vec{F}_\mathrm{A}$, $\vec{F}_\mathrm{B}$ are forces exerted on the particle by the surrounding liquid and the solid wall, respectively. Here we have normalized the mass to be 1. In this formulation, we use the symbol $\vec{F}_\mathrm{B}$, but the reader should be aware that its physical meaning is the same as of the symbol $F$ used in the main text (here we are dealing with the generalized 3 dimensional case while the equations in the main text is one dimensional).
At $t=0$, the interfacial particle of interest $i$ is at position $\vec{r}_i$ and has velocity $\vec{u}_i$.
We now introduce projection operators $\mathrm{\bf{P}}=|\vec{u}_i\rangle\langle \vec{u}_i|\langle \vec{u}_i\vec{u}_i\rangle^{-1}$ and $\mathrm{\bf{Q}}=1-\mathrm{\bf{P}}$, so that
\begin{equation}
\mathrm{e}^{\mathrm{i\bf{L}}t}=\mathrm{e}^{\mathrm{i\bf{QL}}t}+\int_0^t\mathrm{e}^{\mathrm{i}\mathrm{\bf{L}}(t-t')}\mathrm{i\bf{PL}}\mathrm{e}^{\mathrm{i}\mathrm{\bf{QL}}t'}\,\mathrm{d}t',
\label{Dyson}
\end{equation}
where the inner product between two vectors $a, b$ in the Hilbert space $\langle ab\rangle$ is defined as the ensemble average at equilibrium. Since we are interested in the interfacial particle $i$, we only average the position $\vec{r}_i$ over the interfacial region we defined. In a three dimensional system in space, infinite dimensional vectore $a$ and $b$ in the Hilbert space are also three dimensional vectors in space, and $\langle ab\rangle$ is therefore a tensor in space.

Using the operator identity (Eq.~(\ref{Dyson})), we can rewrite Eq.~(\ref{Force decomposition}) as
\begin{equation}
\begin{split}
\dot{\vec{u}}_i(t)=-\frac{1}{kT}\int_0^t\vec{u}_i(t-t')\langle \mathrm{i\bf{L}}\vec{u}_i\mathrm{e}^{\mathrm{i}\mathrm{QL}t'}\mathrm{i\bf{A}}\vec{u}_i\rangle\,\mathrm{d}t'\\+\mathrm{e}^{\mathrm{i\bf{QL}}t}\mathrm{i\bf{A}}\vec{u}_i+\mathrm{e}^{\mathrm{i\bf{L}}t}\mathrm{i\bf{B}}\vec{u}_i.
\end{split}
\label{projection}
\end{equation}
We define $\vec{R}(t)=\mathrm{e}^{\mathrm{i\bf{QL}}t}\mathrm{i\bf{A}}\vec{u}_i$ to be the random force. Then $kT\boldsymbol{\gamma}(t)=\langle \mathrm{i\bf{L}}\vec{u}_i\mathrm{e}^{\mathrm{i\bf{QL}}t}\mathrm{i\bf{A}}\vec{u}_i\rangle=\langle \vec{R}(0)\vec{R}(t) \rangle+\langle \vec{F}(0)\vec{R}(t) \rangle$ defines the memory kernel tensor. 
More explicitly, 
\begin{equation}
\langle \vec{R}(0)\vec{R}(t) \rangle=\langle \mathrm{i\bf{A}}\vec{u}_i\mathrm{e}^{\mathrm{i\bf{QL}}t}\mathrm{i\bf{A}}\vec{u}_i\rangle, 
\end{equation}
\begin{equation}
\langle \vec{F}(0)\vec{R}(t) \rangle=\langle \mathrm{i\bf{B}}\vec{u}_i\mathrm{e}^{\mathrm{i\bf{QL}}t}\mathrm{i\bf{A}}\vec{u}_i\rangle.
\end{equation}
At equilibrium, we have $\langle\vec{u}_i\vec{F}_\mathrm{A} \rangle=\langle\vec{u}_i\vec{F}_\mathrm{B} \rangle=0$ as the result of antisymmetries under time reversal. This means that $\mathrm{|i\bf{A}}\vec{u}_i \rangle$ and $\mathrm{|i\bf{B}}\vec{u}_i \rangle$ are both orthogonal to $|\vec{u}_i \rangle$. Applying the projection, we have $\mathrm{\bf{P}}|\mathrm{i\bf{A}}\vec{u}_i \rangle=\mathrm{\bf{P}}|\mathrm{i\bf{B}}\vec{u}_i \rangle=0$ and $\mathrm{\bf{Q}}|\mathrm{i\bf{A}}\vec{u}_i \rangle=\mathrm{|i\bf{A}}\vec{u}_i\rangle, \mathrm{\bf{Q}}|\mathrm{i\bf{B}}\vec{u}_i \rangle=\mathrm{|i\bf{B}}\vec{u}_i\rangle$. Therefore the operator $\mathrm{\bf{QL}}$ is Hermitian in the subspace of $\mathrm{|i\bf{A}}\vec{u}_i \rangle$ and $\mathrm{|i\bf{B}}\vec{u}_i \rangle$.

If $\langle \vec{F}(0)\vec{R}(t) \rangle=0$, then $kT\boldsymbol{\gamma}(t)=\langle\vec{R}(0)\vec{R}(t) \rangle$ recovers the fluctuation-dissipation theorem. This assumption is not formal under all conditions. In fact, whether $\langle \vec{F}(0)\vec{R}(t) \rangle$ equals to zero or not depends on the properties of $\vec{F}_\mathrm{A}$ and $\vec{F}_\mathrm{B}$ or, equivalently, on the properties of the  operators $\mathrm{\bf{A}}$ and $\mathrm{\bf{B}}$. In our case, we show that every element of $\langle \vec{F}(0)\vec{R}(t) \rangle$ is zero except $\langle F_z(0)R_z(t) \rangle$ ($z$ is the norm direction of the wall) in the following way. Let us take the time derivative
\begin{equation}
\begin{split}
\frac{\mathrm{d}}{\mathrm{d}t}\langle \vec{F}(0)\vec{R}(t) \rangle&=\frac{\mathrm{d}}{\mathrm{d}t}\langle \mathrm{i\bf{B}}\vec{u}_i\mathrm{e}^{\mathrm{i\bf{QL}}t}\mathrm{i\bf{A}}\vec{u}_i\rangle\\
&=\langle \mathrm{i\bf{B}}\vec{u}_i\mathrm{i\bf{QL}}\mathrm{e}^{\mathrm{i\bf{QL}}t}\mathrm{i\bf{A}}\vec{u}_i\rangle\\
&=-\langle \mathrm{i\bf{LQ}}\mathrm{i\bf{B}}\vec{u}_i\mathrm{e}^{\mathrm{i\bf{QL}}t}\mathrm{i\bf{A}}\vec{u}_i\rangle\\
&=-\langle \mathrm{i\bf{L}}\mathrm{i\bf{B}}\vec{u}_i\mathrm{e}^{\mathrm{i\bf{QL}}t}\mathrm{i\bf{A}}\vec{u}_i\rangle\\
&=-\langle \mathrm{i\bf{L}}\vec{F}_{\mathrm{B}}(0)\mathrm{e}^{\mathrm{i\bf{QL}}t}\mathrm{i\bf{A}}\vec{u}_i\rangle\\
&=-\langle (\nabla\vec{F}_{\mathrm{B}}\cdot \vec{u}_i)\mathrm{e}^{\mathrm{i\bf{QL}}t}\mathrm{i\bf{A}}\vec{u}_i\rangle\\
&=-\langle[\nabla\vec{F}_{\mathrm{B}}][\vec{u}_i\mathrm{e}^{\mathrm{i\bf{QL}}t}\mathrm{i\bf{A}}\vec{u}_i]\rangle,
\label{dt}
\end{split}
\end{equation}
where $\nabla\vec{F}_{\mathrm{B}}$ and $\vec{u}_i\mathrm{e}^{\mathrm{i\bf{QL}}t}\mathrm{i\bf{A}}\vec{u}_i$ are tensors.

Since $\vec{F}_\mathrm{B}$ depends only on the position of the particle of interest $\vec{r}_i$ and $\mathrm{i\bf{A}}\vec{u}_i$ only on $\vec{r}_j-\vec{r}_i$ ($j\neq i$), we can take the ensemble average by first integrating over the subspace $\Gamma_{\vec{r}_i}=(\vec{r}_1,\vec{r}_2,...,\vec{r}_{i-1},\vec{r}_{i+1},...,\vec{r}_N,\vec{p}_1,\vec{p}_2,...,\vec{p}_N)$, then integrating over $\vec{r}_i$. We introduce the following abbreviation for simplicity:
\begin{equation}
\langle ab\rangle_{\vec{r}_i}=\int\mathrm{d}\Gamma_{\vec{r}_i} f_{\vec{r}_i}(\Gamma_{\vec{r}_i})a(\Gamma_{\vec{r}_i})b(\Gamma_{\vec{r}_i}),
\label{inner}
\end{equation}
where $f_{\vec{r}_i}$ is the equilibrium distribution of the system when particle $i$ is fixed at $\vec{r}_i$. Thus the time derivitive reforms as
\begin{equation}
\frac{\mathrm{d}}{\mathrm{d}t}\langle \vec{F}(0)\vec{R}(t) \rangle=
-\int\vec{r}_if(\vec{r}_i)\vec{[\nabla}\vec{F}_{\mathrm{B}}(\vec{r}_i)]\langle\vec{u}_i\mathrm{e}^{\mathrm{i\bf{QL}}t}\mathrm{i\bf{A}}\vec{u}_i\rangle_{\vec{r}_i}.
\end{equation}

If we define $\mathrm{\bf{P}}_{\vec{r}_i}|...\rangle=\langle \vec{u}_i...\rangle_{\vec{r}_i}\langle \vec{u}_i\vec{u}_i\rangle_{\vec{r}_i}^{-1}|\vec{u}_i\rangle_{\vec{r}_i}$, where $|\vec{u}_i\rangle_{\vec{r}_i}$ is the subspace of $|\vec{u}_i\rangle$ with $\vec{r}_i$ fixed, we will have $\langle \vec{u}_i\mathrm{e}^{\mathrm{i\bf{Q}}_{\vec{r}_i}\mathrm{\bf{L}}t}\mathrm{i\bf{A}}\vec{u}_i\rangle_{\vec{r}_i}=0$ where $\mathrm{\bf{Q}}_{\vec{r}_i}=1-\mathrm{\bf{P}}_{\vec{r}_i}$. 
This is because $\mathrm{\bf{P}}_{\vec{r}_{i}}|\vec{u}_i\rangle_{\vec{r}_{i}}=|\vec{u}_i\rangle_{\vec{r}_{i}}$ and $\mathrm{\bf{Q}}_{\vec{r}_{i}}|\vec{u}_i\rangle_{\vec{r}_{i}}=0$. 
Due to the property of the operator $\mathrm{\bf{Q}}_{\vec{r}_{i}}$, $\mathrm{e}^{\mathrm{i\bf{Q}}_{\vec{r}_{i}}\mathrm{i\bf{L}}t}$ will keep the vector it operates on to be orthogonal with $|\vec{u}_i \rangle_{\vec{r}_{i}}$ if the vector is orthogonal with $|\vec{u}_i \rangle_{\vec{r}_{i}}$ at time zero, which is true for $|\mathrm{i\bf{A}}\vec{u}_i \rangle_{\vec{r}_{i}}$.
Since $\mathrm{\bf{P}}_{\vec{r}_{i1}}|\vec{u}_i\rangle_{\vec{r}_{i2}}=\delta(\vec{r}_{i1}-\vec{r}_{i2})|\vec{u}_i\rangle_{\vec{r}_{i2}}$ and $\mathrm{\bf{P}}=\sum_{\vec{r}_i}\mathrm{\bf{P}}_{\vec{r}_i}$, we have $\mathrm{\bf{P}}|\vec{u}_i\rangle_{\vec{r}_i}=|\vec{u}_i\rangle_{\vec{r}_i}$. 
Based on the projection property of  $\mathrm{\bf{P}}$ on $|\vec{u}_i\rangle_{\vec{r}_i}$, one gets $\langle \vec{u}_i\mathrm{e}^{\mathrm{i\bf{Q}}\mathrm{\bf{L}}t}\mathrm{i\bf{A}}\vec{u}_i\rangle_{\vec{r}_i}=0$, which means that $\langle \vec{F}(0)\vec{R}(t) \rangle$ would remain constant as $t$ goes to infinity. As in the infinite time limit $\vec{F}(0)$ is not correlated with $\vec{R}(t)$, $\langle \vec{F}(0)\vec{R}(t) \rangle=\langle\vec{F}\rangle\langle\vec{R}\rangle$.
At equilibrium, the net force $\langle F_x\rangle$ ($\langle R_x\rangle$) and $\langle F_y\rangle$ ($\langle R_y\rangle$) that are parallel to the solid wall vanish, but $\langle F_z\rangle$ ($\langle R_z\rangle$) in general does not. Therefore except $\langle F_z(0)R_z(t) \rangle$, all the elements of $\langle \vec{F}(0)\vec{R}(t) \rangle$ are zero. Due to spatial symmetry under reversals in $x$ and $y$ axes, one can show that $\langle R_x(0)R_z(t) \rangle=\langle R_z(0)R_x(t) \rangle=0$ and $\langle R_y(0)R_z(t) \rangle=\langle R_z(0)R_y(t) \rangle=0$. As a result, one can decouple the motion equation norm direction $z$ from the one parallel to the wall. 

In summary, rearranging Eq.~(\ref{Force decomposition}) using the projection operator we defined, we finally have an equation with the form of extended GLE:
\begin{equation}
\begin{split}
\dot{\vec{u}}_i(t)=-\frac{1}{kT}\int_0^t\vec{u}_i(t-t')\langle \mathrm{i\bf{A}}\vec{u}_i\mathrm{e}^{\mathrm{i\bf{QL}}t'}\mathrm{i\bf{A}}\vec{u}_i\rangle\,\mathrm{d}t'\\+\mathrm{e}^{\mathrm{i\bf{QL}}t}\mathrm{i\bf{A}}\vec{u}_i+\mathrm{i\bf{B}}\vec{u}_i(t),
\end{split}
\label{EGLE}
\end{equation}
with the fluctuation-dissipation relation generalized to be
\begin{equation}
\begin{split}
kT\boldsymbol{\gamma}(t)=\langle \mathrm{i\bf{A}}\vec{u}_i\mathrm{e}^{\mathrm{i\bf{QL}}t}\mathrm{i\bf{A}}\vec{u}_i\rangle
=\langle \vec{R}(0)\vec{R}(t) \rangle,
\end{split}
\label{EFD}
\end{equation}
and Eqs.~({\ref{A1}, \ref{A2}}) generalized to be
\begin{equation}
\langle \vec{u}_i(0)\vec{R}_i(t)\rangle=\langle \vec{F}_i(0)\vec{R}_i(t)\rangle=0, \;t>0.
\label{UC}
\end{equation}
One should be aware that the vectors and tensors in Eqs.~(\ref{EGLE}, \ref{EFD}, \ref{UC}) are two dimensional and parallel to the solid wall.
\makeatletter
\setlength{\@fptop}{0pt}
\makeatother

\clearpage


\bibliography{reference}

\end{document}